\documentclass{epl}
 
\title{Evolving Social Weighted Networks: Nonlocal Dynamics of Open Source Communities}
\author{Sergi Valverde\inst{1} \and Ricard V. Sol\'e\inst{1,2}}
\institute{
  \inst{1} ICREA-Complex Systems Lab - 
Universitat Pompeu Fabra, Dr. Aiguader 80, 08003 Barcelona, Spain \\
  \inst{2} Santa Fe Institute - 1399 Hyde Park Road, New Mexico 87501, USA
}


\begin{document}

\maketitle

\begin{abstract}
Complex networks emerge under different conditions through simple rules of growth and evolution. 
Such rules are typically local when dealing with biological systems and most social webs. An important 
deviation from such scenario is provided by communities of agents engaged in technology development, 
such as open source (OS) communities. Here we analyze their network structure, showing that it defines 
a complex weighted network with scaling laws at different levels, as measured by looking at e-mail 
exchanges. We also present a simple model of network growth involving non-local rules based on 
betweenness centrality. The model outcomes fit very well the observed scaling laws, suggesting that 
the overall goals of the community and the underlying hierarchical organization play a key role 
is shaping its dynamics.
\end{abstract}

%---------------------------------------------------------------------------

\section{Introduction}

%---------------------------------------------------------------------------    

Networks predate complexity, from biology to society and technology \cite{Dorogovtsev}. 
In many cases, large-scale, system-level properties emerge 
from local interactions among network components. This is consistent with the general lack 
of global goals that pervade cellular webs or acquaintance networks. However, when 
dealing with large-scale technological designs, the situation can be rather 
different. This is particularly true for some communities of designers working 
together in a decentralized manner. Open source communities, in particular, provide 
the most interesting example, where software is developed through distributed 
cooperation among many agents. The software systems are themselves complex networks [2,3,4],
%\cite{Europhys},\cite{PREMotifs},\cite{Lognets}). 
which have been shown to display small world and scale-free architecture. 
In this paper we analyse the global organization of these problem-solving communities and the possible rules of 
self-organization that drive their evolution as weighted networks.

\begin{figure}
\onefigure[width=0.7\textwidth]{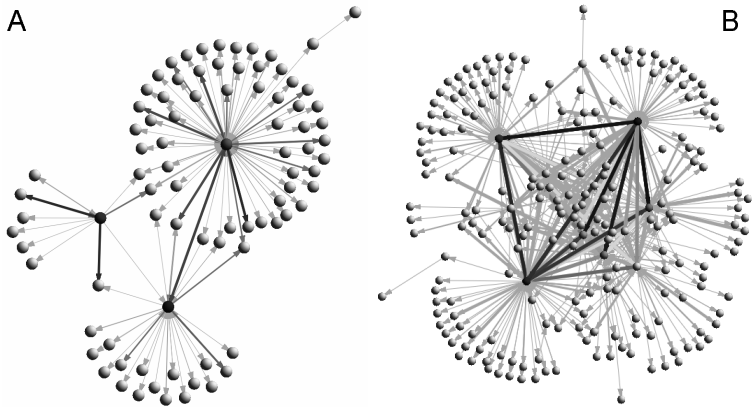}
\caption{ Social network of e-mail exchanges in  open source communities display hierarchical
features. Line thickness represents the number of e-mails flowing from the sender to the receiver. 
Darker nodes and links denote active members and frequent communication,
respectively. (A) Social network for the Amavis community has $N=98$ members, where
the three center nodes display the largest traffic loads. (B) Social network for the TCL community 
has $N=215$ members and average degree $\langle k \rangle \approx 3$.  There is
 a small subgraph of core members (i.e., the hierarchical backbone) concentrating the bulk of 
 e-mail traffic.  Note how strong edges connect nodes with heavy traffic load.}
\label{fig1}
\end{figure}

Following \cite{Crowston2005}, we have analyzed the structure and modeled the evolution
of social interaction in OS communities \cite{Raymond}. Here 
e-mail is an important vehicle of communication and we can recover social interactions 
by analyzing the full register of e-mails exchanged between community members. From this dataset, 
we have focused on the subset of e-mails describing new software errors (bug tracking) and 
in the subsequent e-mail discussion on how to 
solve the error (bug fixing).  Nodes $v_i \in V$ in the social network $G=(V,L)$ represent community 
members while directed links $(i,j) \in L$ denote e-mail communication whether the member $i$ 
replies to the member $j$. At time $t$, a member $v_i$ discovers a new software error (bug) and 
sends a notification e-mail. Then,  other members investigate the origin of  the software bug 
and eventually reply to the message, either explaining the solution or asking for more 
information. Here $E_{ij}(t)=1$ if developer $i$ replies to developer $j$ at time 
$t$ and is zero otherwise. Link weight $e_{ij}$ is the total amount of e-mail traffic 
flowing from developer $i$ to developer $j$:

\begin{equation}
e_{ij}  = \sum\limits_{t = 0}^T {E_{ij} (t)} 
\label{eq:weight}
\end{equation}
where $T$ is the timespan of software development. We have found that
e-mail traffic is highly symmetric, i. e. $e_{ij} \approx e_{ji}$. In order to measure link symmetry, 
we introduce a weighted measure of link reciprocity \cite{recip} namely the {\em link weight reciprocity}
$\rho^w$, defined as

\begin{equation}
\rho ^w  = \frac{{\sum\nolimits_{i \ne j} {(e_{ij}  - \bar e} )(e_{ji}  - \bar e)}}{{\sum\nolimits_{i \ne j} {(e_{ij}  - \bar e} )^2 }}
\label{eq:recipw}
\end{equation}
where $\bar e = \sum\nolimits_{i \ne j} {e_{ij} /N(N - 1)}$ is the average link weight.  This coefficient
enables us to differentiate between weighted reciprocal networks ($\rho^w > 0$) and weighted 
antireciprocal networks ($\rho^w < 0$). The neutral case is given by $\rho^w \approx 0$. 
All systems analyzed here display strong symmetry, with $\rho^w \approx 1$. 
This pattern can be explained in terms of {\em fair reciprocity} \cite{Caldarelli},
where any member replies to every received e-mail.

In the following, we will focus in the analysis of the undirected (and weighted) graph. Let us define 
edge weight (interaction strength) as $w_{ij} = e_{ij} + e_{ji}$, which provides a measure 
of traffic exchanges between any pair of members. Two measures of node centrality are 
frequenly used to evaluate node importance. A global centrality measure is betweeness 
centrality $b_i$ \cite{BC} (i. e. node load\cite{Load}) measured as the number of 
shortests paths passing through the node $i$. Node strength \cite{Barrat2004} is a local measure 
defined as 

\begin{equation}
s_i  = \sum\nolimits_j {w_{ij}}
\label{eq:stren}
\end{equation}
i. e. the total number of messages exchanged between node $i$ and the rest of the community. 
The correlation of centrality measures with local measures (such as undirected degree $k_i$) 
can be used to asses the impact of global forces on  network dynamics.

\begin{figure}
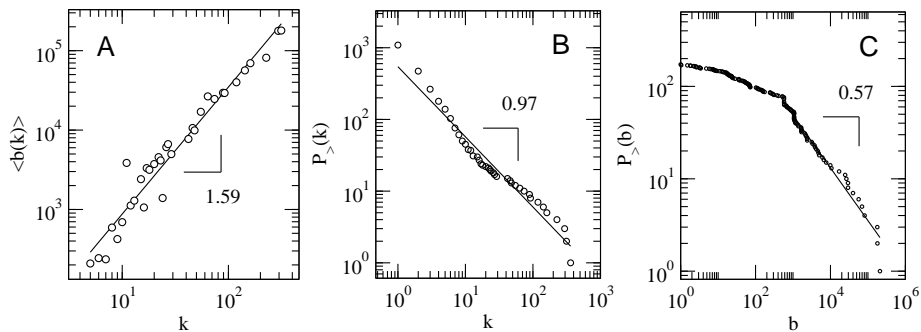

\onefigure[width=0.85\textwidth]{python.load.eps}
\caption{(A) Average betweeness centrality vs degree $\langle b(k)\rangle \sim k^{\eta}$ 
where $\eta \approx 1.59$ for the Python OS community. This exponent is 
close to the theoretical prediction $\eta_{BA} \approx (\gamma -1)/(\delta-1) = 1.70$ (see text). 
(B) Cumulative distribution of undirected degree 
$P_>(k) \sim k^{-\gamma+1}$ with $\gamma \approx 1.97$. (C) Cumulative 
distribution of betweeness centrality $P_>(b) \sim b^{-\delta+1}$ with $\delta \approx 1.57$.}
\label{fig2}
\end{figure}

\begin{table}
\label{t.2}
\begin{center}
\begin{tabular}{|l|c|c|c|c|c|c|c|}
\hline
Project &  $N$  &  $L$ & $\langle k \rangle$ &  $\gamma$ &  $\delta$ & $\eta$ & $(\gamma-1)/(\delta-1)$ \\\hline\hline
Python  & 1090  & 3207  & 2.94 & 1.97 & 1.57  & 1.59 &  1.70 \\\hline
Gaim &  1415 &  2692 &  1.9 &   1.97 &  1.8  &  1.24 &  1.21 \\\hline
Slashcode  &    643 &   1093 &  1.69 &  1.88 &  1.58 &  1.42 &  1.51 \\\hline
PCGEN & 579  &  1654 &  2.85 &  2.04 &  1.67 &  1.54 &  1.55 \\\hline
TCL  &  215  &  590  &  2.74 &  1.97 &  1.33 &  2.34 &  2.93 \\\hline \hline
\end{tabular}
\vspace{0.2 cm}
\caption{Topological measures performed over large OS weighted nets. 
The two last columns at left compare the observed $\eta$ exponent with 
the theoretical prediction $\eta = (\gamma-1)/(\delta-1)$ (see text).}
\end{center}
\end{table}

Figure \ref{fig1} shows two social networks recovered with the above method. 
We can appreciate an heterogeneous pattern of e-mail interaction, where a few nodes 
generate a large fraction of e-mail replies. The undirected degree distribution is a power-law 
$P(k) \sim k^{-\gamma}$ with $\gamma \approx 2$ (see fig. \ref{fig2}B). These social networks 
exhibit a clear hump for large degrees.  Betweeness centrality displays a long tail $P(b)
\sim b^{-\delta}$  with an exponent $\delta$ between 1.3 and 1.8 (see table I and also fig. \ref{fig2}C). 
It was shown that betweeness centrality scales with degree in the network of Internet autonomous 
systems and in the Barab\'asi-Albert network \cite{Barabasi}, as $b(k) \sim k^{-\eta}$. 
From the cumulative degree distribution, i. e. 

\begin{equation}
P_>(k)=\int_k^{\infty}P(k)dk \sim k^{1-\gamma} 
\end{equation}

and the corresponding integrated betweenness, 
with $P_>(b) \sim b^{1-\delta}$, it follows that $\eta = (\gamma-1)/(\delta-1)$ \cite{Vazquez2002}.
The social networks studied here display a similar scaling law with an exponent $\eta$ 
slightly departing from the theoretical prediction (see fig. \ref{fig2}A and table I).
However, a detailed analysis reveals a number of particular features intrinsic to these social networks.

%---------------------------------------------------------------------------

\section{Correlations and Hierarchy in OS Networks}

%---------------------------------------------------------------------------

\begin{figure}
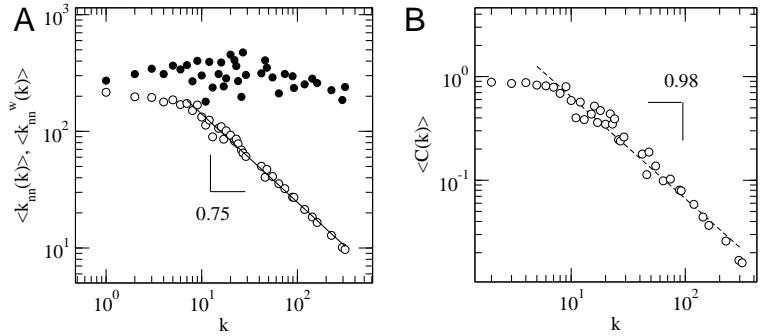

\onefigure[width=0.7\textwidth]{python.corr.eps}
\caption{Correlations in the Python OS community. (A) Average degree of
nearest neighbors vs degree $\langle k_{nn} \rangle \sim k^{\theta}$ where $\theta \approx 0.75$ 
(open circles). The social network is dissasortative from the structural point of view.  However, the 
 weighted average nearest neighbors degree (solid circles) captures more precisely the level of affinity in 
 the community (see text). Instead, traffic is redirected to the core subset of highly 
connected nodes (backbone). (B)  Average clustering vs degree $\langle C(k) \rangle \sim k^{\beta}$ 
with $\beta \approx 1$.}
\label{fig3}
\end{figure}

A remarkable feature of software communities is their hierarchical structure (see fig.\ref{fig1}), 
which introduces non-trivial correlations in the social network topology. We can detect 
the presence of node-node correlations by measuring the average nearest-neighbors degree
$k_{nn} (k) = \sum\nolimits_{k'} {k'P(k|k')}$ where $P(k|k')$ is the conditional probability of
having a link attached to nodes with degree $k$ and $k'$. In the absence of correlations, 
$P(k|k')$ is constant and does not depend on $k$. Here, the average nearest-neighbors degree decays 
as a power-law of degree, $\langle k_{nn} \rangle \sim k^{-\theta}$ with $\theta \approx 0.75$  
(see fig. \ref{fig3}A ). This decreasing behaviour of $k_{nn}$ denotes that low-connected nodes 
are linked to highly connected nodes (see fig. \ref{fig1}A) and thus, these networks are
dissasortative from the topological point of view.  However, the same networks are assortative
when we analyze edge weights. We have observed that frequent e-mail exchanges take place 
between highly connected members.  Following \cite{Barrat2004}, we define the 
weighted average nearest-neighbors degree,

\begin{equation}
k_{nn,i}^w  = \frac{1}{{s_i }}\sum\limits_{j = 1}^k {w_{ij} k_j } 
\label{eq:wknn}
\end{equation}

where neighbor degree $k_j$ is weighted by the ratio $(w_{ij}/s_i)$. According to this definition, 
$k_{nn,i}^w > k_{nn}$ if strong edges point to neighbors with large degree 
and $k_{nn,i}^w < k_{nn}$ otherwise. In software communities,  weighted average 
nearest-neighbors degree is almost uncorrelated with node degree, that is,  $k_{nn,i} \approx $ 
constant (see fig.\ref{fig3}A).  Low connected nodes have weak edges because $k_{nn,i}^w(k)$ is 
only slightly larger than $k_{nn}(k)$ for small $k$ (see fig.\ref{fig3}A). The 
social network is assortative because strong edges attach to nodes with many links, 
i.e., the  difference $k_{nn,i}^w(k) - k_{nn}(k)$ is always positive and increases with degree $k$.
The hierarchical nature of these graphs is well illustrated from the scaling exhibited by the 
clustering $C(k)$ against $k$, which scales as $C(k) \sim 1/k$ (see fig. \ref{fig3}B), consistently 
with theoretical predictions \cite{Clustering}.

%---------------------------------------------------------------------------

\section{Nonlocal Evolution of OS Networks}

%---------------------------------------------------------------------------

\begin{figure}
\onefigure[width=1.0\textwidth]{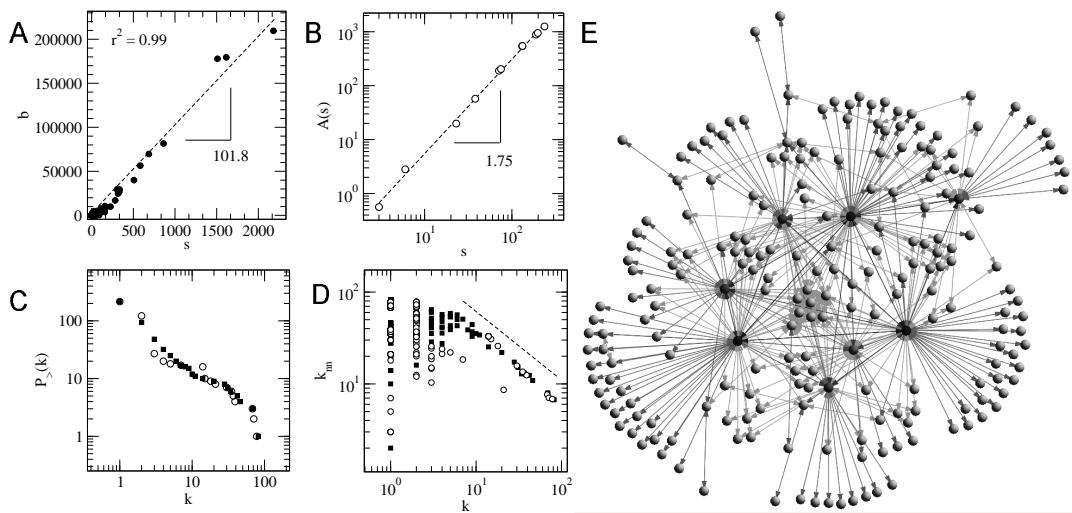}
\caption{ Social network simulation (A) Linear correlation between node strength $s_i$ and  
betweeness centrality (or node load) $b_i$ in the Python community. The correlation coefficient is 0.99. 
This trend has been observed in all communities studied here (B) Estimation of $\alpha$ in the 
TCL project (see text). (C) Cumulative degree distribution in the simulated 
network (open circles) and in the real community (closed squares). All parameters estimated from real data: 
$N= 215$,  $m0 = 15$,  $\langle m \rangle = 3$ and $\alpha =0.75$. Notice the remarkable agreement
between simulation and the real social network. (D) Scaling of average neighbors degree
vs degree in the simulated network (open circles) and in the real social network 
(closed squares). Notice the remarkable overlap of simulation and real data for large $k$. 
(E) Rendering of the simulated network to be compared with the social network displayed in fig. 1B.}
\label{fig:kernel}
\end{figure}

A very simple model predicts the network dynamics of software 
communities, including the shape of the undirected degree distribution $P(k)$ 
and local correlations (see fig.\ref{fig:kernel}C, fig.\ref{fig:kernel}D, and fig.\ref{fig:kernel}E).
The system starts (as in real OS systems) from a fully-connected network of $m_0$ members. 
At each time step, a new member joins the community and a new node is added to
the social network. The new member reports a small number of $m$ e-mails (describing
new software bugs). These new e-mails will be eventually replied by expert community 
members. Member experience is estimated with node strength $s_i$ or 
the total number of messages  sent (and received) by the member $i$ (eq. (\ref{eq:stren})). 
In addition, any member takes into account 
all previous communications regarding any particular software bug.
This suggests that node strength is determined in a nonlocal manner \cite{Goh2005}. 
Indeed, we observe a linear correlation between strength $s_i$ and 
betweeness centrality $b_i$ in software communities (see fig. \ref{fig:kernel}A). 
The probability that individual $i$ replies to the new nember is proportional
to the node load $b_i$,

\begin{equation}
\Pi \left[ {b_i (t)} \right] = \frac{{\left( {b_i (t) + c} \right)^\alpha  }}{{\sum\limits_j {\left( {b_j (t) + c} \right)^\alpha  } }}
\label{eq:kernel}
\end{equation}
where $c$ is a constant (in our experiments, $c=1$) and node load $b_i$ is recalculated 
before attaching the new link, that is, before evaluating eq. (\ref{eq:kernel}).
%Recalculation is an important rule in order to fit real measurements (see below). 
A similar model was presented in \cite{Goh2005}, where 
$b_i$ is recalculated only after the addition of the new node and its $\langle m \rangle$ links. 
Here, the recalculation of betweenness centralities represents a global process of 
information diffusion. Once the target node $i$ is selected, we place a new edge linking 
node $i$ and the new node.

The networks generated with the previous model are remarkably similar to
real OS networks. For example, fig.\ref{fig:kernel} compares our model with the social 
network of TCL software community. The target social network has $N=215$ members 
and $m=\langle k \rangle \approx 3$. A simple modification to a known algorithm for 
measuring preferential attachment in evolving networks \cite{jeong2003} 
enables us to estimate the exponent $\alpha$ driving the attachement rate of 
new links (described in eq. (\ref{eq:kernel})). Due to limitations in available network data 
we have computed the attachment kernel depending on node strength $s_i$ instead of 
node load $b_i$. In order to measure $\Pi \left[ {s_i (t)} \right]$ we compare two 
network snapshots of the same software community at times $T_0$ and $T_1$ where 
$T_0 < T_1$. Nodes in the $T_0$ and $T_1$ network are called {\em "$T_0$ nodes"} 
and {\em "$T_1$ nodes"}, respectively. When a new $i \in T_1$ node joins the network we 
compute the node strength $s_j$ of the $j \in T_0$ node to which the new node $i$ links. Then,
we estimate the attachment kernel as follows

\begin{equation}
\Pi \left[ {s,T_0 ,T_1 } \right] = \frac{{\sum\limits_{i \in T_1 ,j \in T_0 } {m_{ij} \theta (s - s_j )} }}{{\sum\limits_{j \in T_0 } {\theta (s - s_j )} }}
\label{eq:mkernel}
\end{equation}

where $\theta(z)=1$ if $z=0$ and $\theta(z)=0$ otherwise, and $m_{ij}$ is the 
adjacency matrix of the social network. In order to reduce the impact of noise fluctuations,
we have estimated the $\alpha$ exponent from the cumulative function 

\begin{equation}
A(s) = \int\limits_0^s {\Pi (s)ds}.
\end{equation}

Under the assumption of eq. \ref{eq:kernel} the above function scales with 
node strength,  $A(s) \sim s^{\alpha+1}$. Figure \ref{fig:kernel}B displays the cumulative
function $A(s)$ as measured in the TCL software community with $T_0=2003$ and
$T_1=2004$. In this dataset, the power-law fitting of $A(s)$ predicts an 
exponent $\alpha=0.75$. A similar exponent is observed in other systems (not shown). 
In addition, we have estimated the $\alpha_{BA}$ exponent with a preferential attachment 
kernel, $\Pi(k) \sim k^{\alpha_{BA}}$, as in the original algorithm by Jeong et al. \cite{jeong2003}. 
The evolution  of the social networks cannot be described by a linear preferential 
attachment mechanism because the observed exponent is $\alpha_{BA} > 1.4$ (not shown).

%---------------------------------------------------------------------------

\section{Discussion}

%---------------------------------------------------------------------------

The analysis of correlations in open source communities indicates they are closer to the
Internet and communication networks than to other social networks (e.g.,  the
network of scientific collaborations ). The social networks 
analyzed here are dissasortative from the topological point of view and assortative 
when edge weights are taken into account. A distinguished feature of social networks 
in software communities is a subset of core members acting like the community backbone. 
In these communities, the bulk of e-mail traffic is redirected to the 
strongest members, which are reinforced as the dominant ones. 

We have presented a model that predicts many global and local social network measurements 
of software communities. Interestingly, the model suggests that reinforcement is nonlocal, that is, 
e-mails are not independent of previous e-mails. The conclusions of the present work must 
be contrasted with the local reinforcement mechanism proposed by Caldarelli et. 
al. \cite{Caldarelli}. In their model, any pair of members can increase the strength of 
their link with independence of the global activity. Several features of software communities 
preclude the application of their model. For example, fixing a software bug is a global task 
which requires the coordination of several members in the community. Any e-mail response 
requires to consider all the previous communications regarding the specific subject 
under discussion. In addition, their model does not consider a sparse network structure 
and every individual is connected with everybody else, which is not the case of OS communities. 

We can conceive other alternatives instead of computing betweeness centralities 
in eq.(\ref{eq:kernel}). An interesting approach includes the discrete simulation of e-mails 
tracing shortest paths in the social network, as in some models of internet 
routing \cite{SoleValverde2001}. Packet transport-driven simulations can provide good estimations
of the number of e-mails received by any node. Nevertheless, the present model enables us to explain 
remarkably well the OS network dynamics.  Another extension of the model is the addition of
new links between existing nodes, which can provide better fittings to local correlation measures.
Finally, the current model is a first step towards a theory of collaboration and self-organization
in open source communities. In this context, the techniques and models presented here are
useful tools to understand how social collaboration takes place in distributed environments.

\acknowledgments
We thank our colleague Vincent Anton for useful discussions. This work has been 
supported by grants FIS2004-05422, by the EU within the 6th Framework Program 
under contract 001907 (DELIS) and by the Santa Fe Institute.


\begin{thebibliography}{0}

\bibitem{Dorogovtsev}
        \Name{Dorogovtsev, S. N. \and Mendes, J. F. F.}
        \Book{Evolution of Networks: From Biological Nets to the Internet and WWW}
        \Publ{Oxford Univ. Press, New York}
        \Year{2003}.
	
\bibitem{Europhys}
        \Name{Valverde, S., Ferrer-Cancho, R. \& Sol\'e, R. V.}
        {Europhys. Lett.} 
        \Vol{60}
        \Year{2002}
        \Page{512}.

\bibitem{PREMotifs}
        \Name{Valverde, S. \and Sol\'e, R. V.}
        {Phys. Rev. E}
        \Vol{72}
        {026107}
        \Year{2005}.

\bibitem{Lognets}
        \Name{Valverde, S., and Sol\'e, R. V.}
        {Europhys. Lett.} 
        \Vol{72}
	{5}
        \Year{2005}
        \Page{858--864}.
	
\bibitem{Crowston2005}
        \Name{Crowston, K. and Howison, J.}
        {First Monday}
        \Vol{10}
        {2}
        \Year{2005}.
	
\bibitem{Raymond}
     \Name{Raymond, E. S.}
        {First Monday}
        \Vol{3}
        {3}
        \Year{1998}.
	

\bibitem{recip}
        \Name{Garlaschelli, D., and Loffredo, M. I.}
        {Phys. Rev. Lett.}
        \Vol{93}
        {268701}
        \Year{2004}.
	
\bibitem{Caldarelli}
        \Name{Caldarelli, G., Coccetti, F.,  and de Los Rios, P.}
        {Phys. Rev. Lett.}
        \Vol{70}
        {027102}
        \Year{2004}.	
	
\bibitem{BC}
	\Name{Brandes, U.}
	{Journal of Mathematical Sociology}
	\Vol{25}
	{2}
	\Page{163--177}
	\Year{2001}.
	
\bibitem{Load}
        \Name{Goh, K.-I.,  Kahng, B., and Kim, D.}
        {Phys. Rev. Lett.}
        \Vol{87}
        {278701}
        \Year{2001}.

\bibitem{Barrat2004}
	\Name{Barrat, A., Barth\'elemy, M.,  Pastor-Satorras, R., and Vespignani, A.}
	{Proc. Natl. Acad. Sci. USA}
	\Vol{101}
	\Year{2004}
	\Page{3747}.
	
\bibitem{Barabasi}
        \Name{Barab\'asi, A.-L., and Albert, R.}
        {Science}
        \Vol{286}
        \Year{1999}
        \Page{509}.
	
\bibitem{Vazquez2002}
	\Name{Vazquez, A., Pastor-Satorras,  R., and Vespignani,  A.}
        {Phys. Rev. Lett.}
        \Vol{65}
        {066130}
        \Year{2002}.
	
\bibitem{Clustering}
	\Name{Dorogovtsev, S. N., Goltsev, A. V.,  Mendes, J. F. F.}
	{Phys. Rev. E}
	\Vol{65}
	{066122}
	\Year(2002).
	
\bibitem{Goh2005}
	\Name{Goh, K.-I., Kahng, B., and Kim, D.}
	{Phys. Rev. E.}
	\Vol{72}
	{017103}
	\Year{2005}.
	
\bibitem{jeong2003}
        \Name{Jeong, H., N\'eda, Z., and Barab\'asi, A.-L.}
        {Europhys. Lett.}
        \Vol{61}
        {4}
        \Year{2003}
        \Page{567}.

\bibitem{SoleValverde2001}
	 \Name{Sol\'e, R. V., and Valverde, S.}
        {Physica A}
        \Vol{289}
        \Year{2001}
        \Page{595-695}.
	
\end{thebibliography}
\end{document}